\documentclass[a4paper,fleqn]{cas-dc}

\usepackage[numbers]{natbib}
\usepackage{amsmath,amssymb,mathtools}
\usepackage{graphicx}
\usepackage{enumitem}
\usepackage{caption}
\usepackage{hyperref}
\usepackage{bm}
\usepackage{booktabs}
\usepackage{orcidlink}

\def\tsc#1{\csdef{#1}{\textsc{\lowercase{#1}}\xspace}}
\tsc{WGM}
\tsc{QE}
\tsc{EP}
\tsc{PMS}
\tsc{BEC}
\tsc{DE}

\begin{document}
\let\WriteBookmarks\relax
\def\floatpagepagefraction{1}
\def\textpagefraction{.001}
\shorttitle{The shape of the growth index}
\shortauthors{Z. Sakr \& J. Zheng}

\title [mode = title]{The shape function of the observed growth index}                      

\author[1,2]{Ziad Sakr \orcidlink{0000-0002-4823-3757} }[type=editor,
                        orcid=0000-0002-4823-3757
                        ]
\ead{ziad.sakr@net.usj.edu.lb}

\affiliation[1]{organization={Instituto de Física Teórica UAM-CSIC},
                addressline={Campus de Cantoblanco}, 
                city={Madrid},
                postcode={28049}, 
                country={Spain}}
\affiliation[2]{organization={Faculty of Sciences, Universit\'e St Joseph;},
                city={Beirut},
                country={Lebanon}}  
                
\author[3]{Jinglan Zheng }

\affiliation[3]{organization={Fakultät für Physik, Universität Bielefeld},
                addressline={Postfach 100131}, 
                city={Bielefeld},
                postcode={D-33501}, 
                country={Germany}}                              

\begin{abstract}
The growth index $\gamma$ is a powerful trigger for detecting deviations from $\Lambda$CDM. However, its value is often determined by considering an asymptotic constant value that works for all redshift, or else following a chosen parameterisation. Here we formulate the growth index as function of three quantities that could be directly related to observables in redshift bins, $f\sigma_8(z_i)$, $f(z_i)$ and $H(z_i)$. We determine its value and its derivative at observed nodal center of redshift bins and use the shape function method, after showing insightful connection with its underlying governing virtual-work conservation principle, often used in other disciplines than cosmology, to construct a redshift dependence of the $\gamma$ without assuming a specific parameterization. We then use the resulting growth index shape function to test if we can disentangle between different scenarios where there might be discrepancies between its three constituent measured components. We also tested whether it can be used to rule out models of modified gravity, or extended parametric models of the growth index that capture more general behaviors with an additional parameter as function of the scale factor or dark energy in a flat universe. Adopting forecasted measurements from next generation surveys on the three aforementioned quantities used to construct $\gamma$, we find that reported discrepancies between them could be detected with our method, but at the bins where the errors and lost of precision from our addition of degrees of freedom is small with respect to the deviation of $\gamma$. The same could be concluded for first order extensions to $\gamma$ or common modified gravity models such as $f(R)$ or nDGP, and to a lesser degree with models of dynamical dark energy after supposing the latest DESI reported values. We conclude that this method is a strong tool to investigate cosmology in a model-independent way especially with forthcoming data delivered by further stage-IV surveys with more stringent uncertainties.

%
\end{abstract}

\maketitle

\section{Introduction}

The growth index $\gamma$ provides a compact descriptor of the linear growth of matter density perturbations through the approximation $f(a) \approx \Omega_{\mathrm{m}}(a)^{\gamma}$ \cite{Lahav:1991wc,Wang1998} and is widely used to detect deviations from $\Lambda\mathrm{CDM}$ model such as nonstandard dark energy (time-varying equation of state, clustering) or modifications of gravity that change the effective force law or metric potentials \cite{Polarski2008,Cortes:2024yon,Sakr:2023xnw}. 
Current large-scale structure and RSD measurements, combined with CMB and BAO priors, constrain $\gamma$ at roughly the ten-percent level, with some dataset combinations reporting mild tensions with the $\Lambda\mathrm{CDM}$ expectation \cite{Sakr:2023bms,Specogna2024}. Looking ahead, DESI \cite{DESI:2016igz}, Euclid \cite{Euclid:2024yrr} and Rubin \cite{LSSTScience:2009jmu} are expected to tighten more those constraints and enable more robust tests of GR versus modified gravity.

Common parameterizations include a constant $\gamma$ as a baseline \cite{Linder2005}, two-parameter linear forms such as $\gamma(z) = \gamma_0 +  f(z)\, \gamma_1$ where $f(z)$ could take a first degree simple form as function of the redshift or equivalently the scale factor, or expansions in $1 - \Omega_{\mathrm{m}}(z)$ motivated by analytic series solutions \cite{LinderCahn2007,Polarski2008,Basilakos:2016nyg}, and model-specific expressions derived for $w$CDM, quintessence, $f(R)$, DGP and scalar-tensor theories \cite{Wang1998,LinderCahn2007,Gannouji:2008wt,Steigerwald:2014ava,Wen:2023bcj}. There were also more general but still theoretically motivated parameterizations such as performing a mappings that translates phenomenological modified-gravity function $\mu$ for the Newtonian potential modification (and $\Sigma$ for the Weyl potential) into an effective $\gamma$  (and effective $\gamma_\ell$) \cite{Pogosian2010,Sakr:2023xnw}. 

Besides the previous approaches, one could also consider more data-driven methods, as is our aim in this work, and determine $\gamma$,  avoiding a fixed functional form, through construction and inference methods such as binned $\gamma(z)$, spline reconstructions or Gaussian-process inference. In that regard, we mention \cite{Perenon:2022fgw} that used both expansion and growth observations, in the same direction similar to our approach we present next, to constraint the dark energy equation of state and use it as well to forecast on $\gamma$. We can also mention \cite{Yin:2019rgm}, inferring $\gamma(z)$ from growth and cosmographic parameters from observations and then marginalizing to obtain $\gamma$ only, or \cite{Arjona:2020kco} that went into expressing $\gamma$ from only $f\sigma_8$ measurments. However, its value there at $z=0$ was still needed so that they have to assume, or further obtain $\sigma_8(z=0)$ parameter from other observations. Finally, and more closely related to our approach in this work is \cite{Oliveira:2023uid} one, who expressed $\gamma$ as fully function of the three interesting observables: the growth rate $f(z)$, its product with the matter fluctuation $f\sigma_8$ and the expansion factor $H(z)$. However, they only limited it to the case of $\gamma$ constant. 

In this work, we shall also express $\gamma$ as function of the three latter mentioned observables, which will allow to simultaneously detect deviations from $\Lambda$CDM from models that could modify one or many of the three observables, but also it will allow to detect discrepancies between probes that are independently measuring each of the involved observables. We shall not limit however to the case of a constant $\gamma$ value. Furthermore,  the works above used either Gaussian process reconstruction, or machine learning methods to reconstruct $\gamma$ from values determined at the nodal points or redshift values where the measurements were taken (which could be the $z$ at the center of the bins since data is usually delivered after binning). Here we propose another method, in order to reconstruct $\gamma$, based on a variational principle, through the use of the shape function method \cite{Pierron2012}, which is a mathematical formulation based on the virtual-work conservation principle, used in engineering and physics primarily in the context of finite element analysis and structure loads calculations, as it helps in interpolating the values of a function within a finite element by providing a relationship between the degrees of freedom (nodal values) and the field variables (such as the redshift or scale factor in our case) within the element.

This work is organized as follows. After introducing the subject, we next, in Sect.~\ref{sect:gamtheo}, present the main equations of the expression of $\gamma$ in the chosen observables. We also present the shape function and the formalism relating it to our $\gamma(z)$ reconstruction. Then, in Sect.~\ref{sect:TestandForcast}, we present the datasets and the cases and results from forecasys performed using our methodology, before we conclude in Sect.~\ref{sect:concl}.

\section{Formalism and reconstruction of $\gamma(a)$}\label{sect:gamtheo}

\subsection{The growth index expression from observables}

As mentioned in the introduction, we aim at expressing $\gamma(z)$ as function of $f\sigma_8(z)$, $f(z)$ and $H(z)$\footnote{We caution the reader and note that the quantity that enters the equations, or that we obtain from the measurements, is rather $E(z)$, the normalized Hubble parameter, but we will use interchangeably $H(z)=H_0\, E(z)$ instead throughout the article.}. 
We start from the generalization of the usual formula for the growth index where we consider that it is dependent of a function of redshift (or scale factor)  
\begin{equation}\label{eq:gam}
f = \Omega(a)^{\gamma\left( {\rm f}(a) \right)}
\end{equation}
which, after choosing a modification that allows to introduce  $H(a)$ and $\gamma'$ can be written as (after dropping, to simplify the  display, the scale factor notation for $\gamma$ and all $f$ or subsequent $f \sigma_8$ products)
\begin{equation}\label{eq:devgam}
\frac{f'}{f} = \gamma A(a) + {\gamma}' \frac{\ln f}{\gamma}
\end{equation}
where $A(a) \equiv\frac{2q(a)-1}{a}$ and $q(a)\equiv-1-\frac{a}{H}\frac{dH}{da}$ and prime denotes derivative with respect to $a$. This first step is important because the matter density and its derivative appear in the equation in the second and first term respectively, where we choose to express the first term $\Omega_{\rm m}'/\Omega_{\rm m}$ as a function of $H(z)$ because we consider that the density will change over time following the background expansion, and $\Omega_m$ alone that enters the second term as function of $f$ and $\gamma'$ because here the matter density has the role of a growth quantity. Moreover, the matter density cannot be expressed as function of a distance probe because we need to know $H_0$ for that, unless we combine many probes to break the degeneracy.
On the other hand we also can write the following 
\begin{equation}\label{eq:fs8primeprimefs8}
\frac{(f\sigma_8)''}{f\sigma_8} = \frac{   \frac{f''}{f}  + 2\frac{f'}{f} \frac{\sigma_8'}{\sigma_8} + \frac{\sigma_8''}{\sigma_8}       }{\frac{f'}{f} + \frac{\sigma_8'}{\sigma_8}                        }
\end{equation}
where
\begin{equation}
\frac{\sigma_8'}{\sigma_8} = \frac{f}{a}  \quad {\rm and} \quad \frac{\sigma_8''}{\sigma_8} = \frac{f^2-f}{a} + \frac{f}{a}\left(\frac{\ln f}{\gamma}\gamma'+\gamma A\right)
\end{equation}
and 
\begin{equation}
\frac{f''}{f} = 2 \, \gamma' A(a) + \left( \frac{\ln f}{\gamma} \gamma' +\gamma A(a)  \right) ^2 +\gamma A'(a)
\end{equation}
with $A' = 2 \frac{q'(a)}{a} - \frac{2q(a)-1}{a^2}$  and after limiting our models to first derivatives of the growth index thus considering that $\gamma''$ is zero, it yields at the end, after inserting in Eq.~\ref{eq:fs8primeprimefs8} 
\begin{align}
\frac{(f\sigma_8)''}{f\sigma_8} = & \, 2 \, \left( \frac{(f \sigma_8) '}{f \sigma_8 } - \frac{f}{a} - \gamma A \right) \left( \frac{\gamma A}{\ln f} \right) A {\nonumber}\\
& + \gamma A' + \left( f^2 - f \right) / a^2 {\nonumber}\\
& + 3 \, \frac{f}{a} \left( \frac{(f \sigma_8) '}{f \sigma_8 } - \frac{f}{a} \right) + \left( \frac{f \sigma_8 '}{f \sigma_8 } - \frac{f}{a} \right)^2
\end{align}
We then solve this equation for $\gamma$ analytically to get
\begin{equation}\label{eq:gamma_code}
\gamma(a)=\frac{1}{4}\frac{\Big(A'(a)\,a\,{f \sigma_8} \, {\ln f} + 2\,a\,(f \sigma_8)' - 2\,f\,f \sigma_8 - 2\sqrt{2}\,{\sqrt{\Delta(a)}}\Big)}{a A(a) f \sigma_8}
\end{equation}
where the lengthy expression under the square root is 
\begin{align*}
&\Delta(a) = -A(a)\,a^{2}f \sigma_8\,(f \sigma_8 )''\ln f + A(a)\,a^{2}(f \sigma_8 )'^{2}\ln f \\
&+ A\,a\,f\,f \sigma_8\,(f \sigma_8 )'\ln f - A(a) \, f^{2}(f \sigma_8)^{2}\ln f \\
&- A(a)\,f\,f \sigma_8^{2}\ln f + \tfrac{1}{8}A'(a)^{2}a^{2}(f \sigma_8)^{2}\ln f^{2}\\ 
&+ \tfrac{1}{2}A'(a)\,a^{2}f \sigma_8\,(f \sigma_8 )'\ln f  - \tfrac{1}{2}A'(a)\,a\,f\,(f \sigma_8)^{2}\ln f \\
&+ \tfrac{1}{2}a^{2}(f \sigma_8 )'^{2} - a\,f\,f \sigma_8\,(f \sigma_8 )' + \tfrac{1}{2}f^{2}(f \sigma_8)^{2}
\end{align*}
and then get $\gamma'$ in each node
\begin{equation}\label{eq:gamma_prime_code}
\gamma'(a)=\frac{(f \sigma_8 )'/f \sigma_8 - f/a - \gamma \, A}{\ln f/\gamma(a)}
\end{equation}

\begin{figure*}[h]
\centering
\includegraphics[width=\textwidth]{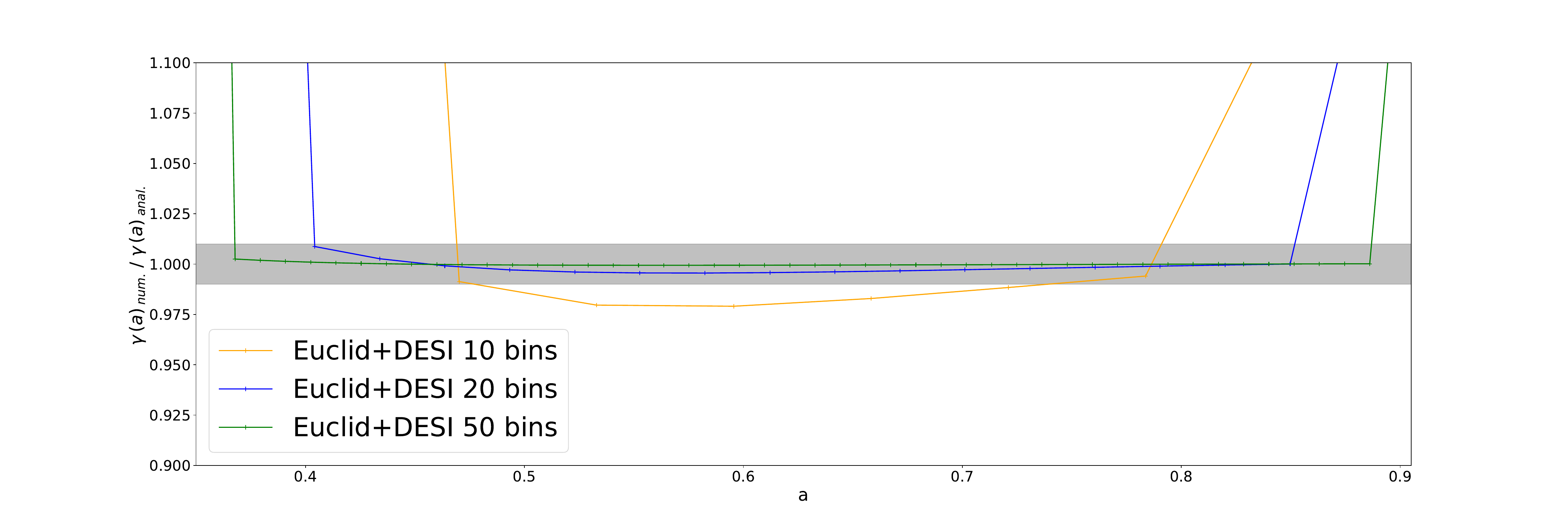}
\caption{A comparison from our constructed $\gamma$ obtained at different bins in comparison to analytic values. The grey band marks the 1\% limit.}
\label{fig:GIgamforecastEuclidDESIresidual}
\end{figure*}

\noindent We show in Fig.~\ref{fig:GIgamforecastEuclidDESIresidual} a comparison from our constructed $\gamma$ obtained at different bins in comparison to the fiducial $\Lambda$CDM values as a null test that could be possible because in this case $\gamma'$ should vanish and unify the `geometric' and 'growth' nature of the matter density. Other more sophisticated models such as ones with an extension to $\gamma$ or models that modifies the expansion or the growth in less trivial ways are more difficult to be alternatively mapped into an effect of the growth index on the observables (hence the utility of our method which we will apply on such models to explore their effects on $\gamma$). The sample redshift bins, or the cosmological parameter values, are the same ones we have used later to forecast and test the ability of our method to distinguish models in Sect.~\ref{sect:TestandForcast}. We checked for three sampling divisions. They are all in broad good agreement but the 1\% threshold is reached for 20 bins, and even more for 50 bins. For later we shall then adopt the 20 bins sampling since 50 bins would be much higher than what is usually adopted for such binning schemes.

\subsection{Variational reconstruction: principle-of-virtual-work shape function}

The shape function is a mathematical formulation, derived from applying the virtual-work conservation principle, that is used in engineering and physics primarily in the context of finite element analysis and other numerical methods. It helps in interpolating the values of a function within a finite element by providing a relationship between the degrees of freedom (nodal values) and the field variables (such as displacement, temperature, etc.) within the element.

To see how it applies for our purposes, where the following, except for the first equation, is proper to this article, we start from the full weak form that translates the virtual-work conservation principle
\[
\delta W_{\text{ext}}=\int_V \sigma\,\delta\varepsilon\,dV+\int_{\partial\Omega} T\,\delta u\,dS+\int_{\partial\Omega} M\,\delta\theta\,dS,
\]
where external applied forces $W_{\text{ext}}$ equates internal stress-strain effects (first term), work from traction-displacement (second term) and Moment-bending work (third term). Then for our function interpolation-by-constraint case we remove any other external tractions and treat internal distributed terms as absent or accounted for separately, leaving the boundary/point traction $T_i$ and moment $M_i$ (its derivative) generalized loads with respectively a virtual displacement $\delta u(x)$ (which will be our $\gamma(x)$ in our case) and $\delta \theta = \delta u'(x)$ .

\noindent Then we use a penalty functional energies to represent continuous values taking into account the nodes fixed ones,
\begin{equation}
\Pi_{\text{pen},i}=\tfrac{1}{2}\alpha_i\bigl(u(x_i)-U_i\bigr)^2+\tfrac{1}{2}\beta_i\bigl(u'(x_i)-\Theta_i\bigr)^2,
\end{equation}
(with $\alpha$ and $\beta$ penalty coefficients) which variation gives the conjugate generalized forces
\begin{equation}
T_i=\alpha_i\bigl(u(x_i)-U_i\bigr),\qquad
M_i=\beta_i\bigl(u'(x_i)-\Theta_i\bigr),
\end{equation}
and the virtual‑work contribution
\begin{align}
T_i\,\delta u(x_i)+M_i\,\delta u'(x_i)
& =  \alpha_i\bigl(u(x_i)-U_i\bigr)\delta u_i \nonumber \\ 
+ \beta_i\bigl(u'(x_i)-\Theta_i\bigr)\delta u'_i.
\end{align}

\noindent which stationarity condition independent of $\alpha$ and $\beta$ 
\begin{equation}
\sum_i\bigl[\alpha_i\bigl(u(x_i)-T_i\bigr)\,\delta u_i
+\beta_i\bigl(u'(x_i)-M'_i\bigr)\,\delta u'_i\bigr]=0,
\end{equation}could be indeed obtained by enforcing $u(x_i)=T_i$ and $u'(x_i)=M_i$.

More practically, we have calculated $\gamma(a)$ and $\gamma'(a)$ at each node following Sect.~\ref{sect:gamtheo} and we want to construct a function $u$ at each scale factor value $a$. We adopt a third order polynomial for $u$ and write the system of equations at each bin imposing the growth index and its derivative values at the border where $u_1=u(a_1)$ and $u_2=u(a_2)$ 

\begin{align}
& x_1 + x_2\,u_1 + x_3\,u_1^2 + x_4\,u_1^3 = \gamma(a_1)
\\
& x_2 + 2\,x_3\,u_1 + 3\,x_4\,u_1^2 = \gamma'(a_1)      
\\
& x_1 + x_2\,u_2 + x_3\,u_2^2 + x_4\,u_2^3 = \gamma(a_2)
\\
& x_2 + 2\,x_3\,u_2 + 3\,x_4\,u_2^2 = \gamma'(a_2)
\end{align}
which we solve for the $x_i$ terms at the bounderies $a_1$ and $a_2$ to obtain:
\begin{align}
x_1  = & \,  (a_1^3 (\gamma(a_2) - a_2 \gamma'(a_2)) + a_1^2 a_2 (a_2 (\gamma'(a_2)
\\& - \gamma'(a_1)) - 3 \gamma(a_2)) + a_1 a_2^2 (a_2 \gamma'(a_1) + 3 \gamma(a_1))  {\nonumber}
\\ & - a_2^3 \gamma(a_1))/(a_1 - a_2)^3 {\nonumber}
\end{align} 
\begin{align}
x_2 =& \, (a_1^3 \gamma'(a_2) + a_1^2 a_2 (2 \gamma'(a_1) + \gamma'(a_2))
\\& - a_1 a_2 (a_2 \gamma'(a_1)+ 2 a_2 \gamma'(a_2) {\nonumber}
\\&+ 6 \gamma(a_1) - 6 \gamma(a_2)) - a_2^3 \gamma'(a_1))/(a_1 - a_2)^3 {\nonumber}
\end{align} 
\begin{align}
x_3 & =  (a_1^2 (-(\gamma'(a_1) + 2 \gamma'(a_2))) + a_1 (-a_2 \gamma'(a_1) 
\\ &+ a_2 \gamma'(a_2)+ 3 \gamma(a_1) - 3 \gamma(a_2))  {\nonumber} 
\\&+ a_2 (2 a_2 \gamma'(a_1) + a_2 \gamma'(a_2)  {\nonumber}
\\ & + 3 \gamma(a_1) - 3 \gamma(a_2)))/(a_1 - a_2)^3 {\nonumber}
\end{align}
\begin{align}
x_4  = &((a_1 - a_2) (\gamma'(a_1) + \gamma'(a_2)) - 2 \gamma(a_1) 
\\&+ 2 \gamma(a_2))/(a_1 - a_2)^3
\end{align}

\begin{figure*}[h]
\centering
\includegraphics[width=\textwidth]{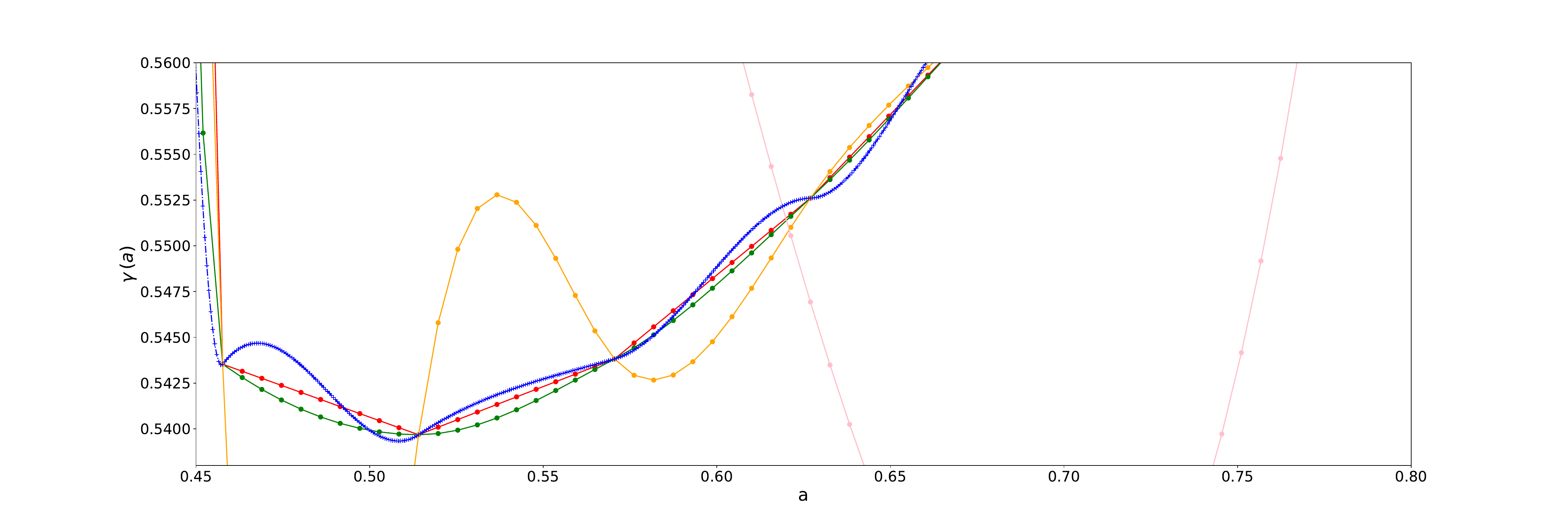}
\caption{Different interpolation methods including simple linear interpolation (red lines), continuity preserving spline interpolation (green lines), B-spline smoothing wise interpolation (pink lines), cubic spline interpolation (orange lines) and this work shape function interpolation (blue lines)}
\label{fig:GIgamw0wainterpcomp}
\end{figure*}

We show in Fig.~\ref{fig:GIgamw0wainterpcomp} different interpolation methods, including ours. The linear interpolation shows abrupt changes while the continuity preserving does not preserve the same for the derivative, the B-spline explore far in between border constraints while the cubic spline one shows smaller but also large deviations between nodes. 

\subsubsection*{Error propagation}
Finally, to propagate the uncertainties measurments on $f(z_i)$, $f\sigma_8(z_i)$ or $H(z_i)$, we approximate using first order formula\footnote{A more elaborated computation would be to include higher order or marginalize from a FIsher matrix computation, task we leave for future refining works on the method}
$\delta\gamma(z) \approx \sqrt{\sum_i\left(\frac{\partial\gamma(z)}{\partial \theta_i}\,\delta \theta_i\right)^2}$, where $theta_i$ stands for the errors for each quantity, or intermediate derived quantities such as $A(z_i)$, $A'(z_i)$ or $f'(z_i)$,   $f'\sigma_8(z_i)$ and  $f''\sigma_8(z_i)$ in each bin, all implemented using finite-difference gradients. We then follow the same for $\delta \gamma'(z)$ in order to apply the same construction method through shape function to simiraly construct the continuous 'profile' of the error bars around the fiducial $\gamma(z)$.  

\section{Tests and forecasts}\label{sect:TestandForcast}

In order to estimate errors for $f(z_i)$ and $H(z_i)$ values, we use and follow \cite{Zheng:2023yco} considering values of $\sigma_f$ and $\sigma_H$ from Table 2 in that work obtained from forecasts from a survey that approximates a full shape power spectrum probe from spectroscopic DESI-like survey at low redshift combined with a Euclid-like one at higher redshift, with specifications according to Table 1 in \cite{Zheng:2023yco}. Since here we need only the constraints on $H(z_i)$ and $f(z_i)$, we selected from the resulting full  parameter covariance matrix  only the corresponding entries for $n_b=10$ redshift bins, with size $\Delta z=0.2$ and central redshifts $\bar{z}=\{$0.1, 0.3, 0.5, 0.7, 0.9, 1.1, 1.3, 1.5, 1.7, 1.9$\}$. These estimations were obtained after cutting at a conservative non-linear scale $k_{\rm max}=0.25 \, h/$Mpc for the one-loop power spectrum.
We still need to obtain errors on $f\sigma_8(z_i)$ which could be obtained from scales at the linear level while to constrain $f$ alone the extension to non-linear power spectra and/or bispectra was necessary as was the case in \cite{Amendola:2023awr} \footnote{Note that, it is not necessary to get $f$ and $fs8$ from the same probe since, as we see later, $fs8$ could come from redshift distortion probes while $f$ could be obtained from integrating a growth obtained from other probes such as galaxy angular correlation functions or galaxy cluster counts. In the same manner, $H(z)$ could be obtained from other probes such Cosmic Chronometers}. 
We thus conducted a Fisher matrix estimation of $f\sigma_8(z_i)$ using the \texttt{Cosmicfish} code we modified to include a $f\sigma_8(z)$ parametrised binning. We stayed at the linear level cutting at $k_{\rm max}=0.25 \, h/$Mpc and adopted a semi-analytic nonlinear RSD model, with free terms for each bin, that also includes Fingers-of-God and BAO damping. We obtain errors for the same bins as \cite{Amendola:2023awr}. Then because we wanted a sampling of 20 bins and not 10 bins we interpolated from 10 to 20 bins using a simple linear method, enough for our purpose, but we increased the errors, supposing a Poisson distribution, by afactor of  $\sqrt{2}$ to account for a statistical lost that would have happened in the errors estimation when observed quantities are re-measured. 

We plan to forecast on different scenarios with the first supposing the fiducial is obtained by different constant values of $\gamma$ or extension with an additional parameter $\gamma_1$; while in the second we consider models of dynamical dark energy or modified gravity models; and the last is supposing we have a discrepancy in either of the three probes. For the latter case, we recompute  $f\sigma_8$, $f(z_i)$ and $H(z_i)$ with different values for $\Omega_{\rm m}$ for each discrepant one; while we compute them for the MG or DE models  using cosmological solver \texttt{MGCLASS} \cite{Sakr:2021ylx}. Finally it remains the subtle calculation of $\sigma_8$ when we consider 'pure' $\gamma$ models. For that power spectrum entering its calculation is rescaled for the new growth factor $D$ obtained after integrating the growth rate $f$ following the new $\gamma$.\\

\begin{figure*}[htbp]
\centering
\includegraphics[width=\textwidth]{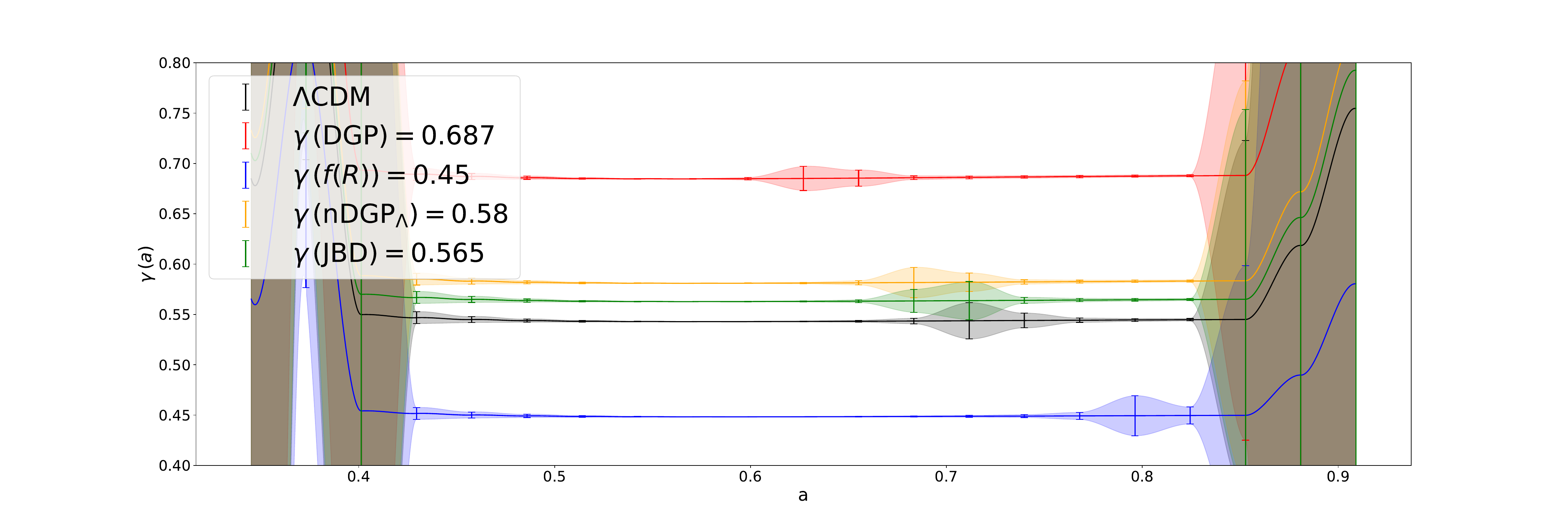}
\includegraphics[width=\textwidth]{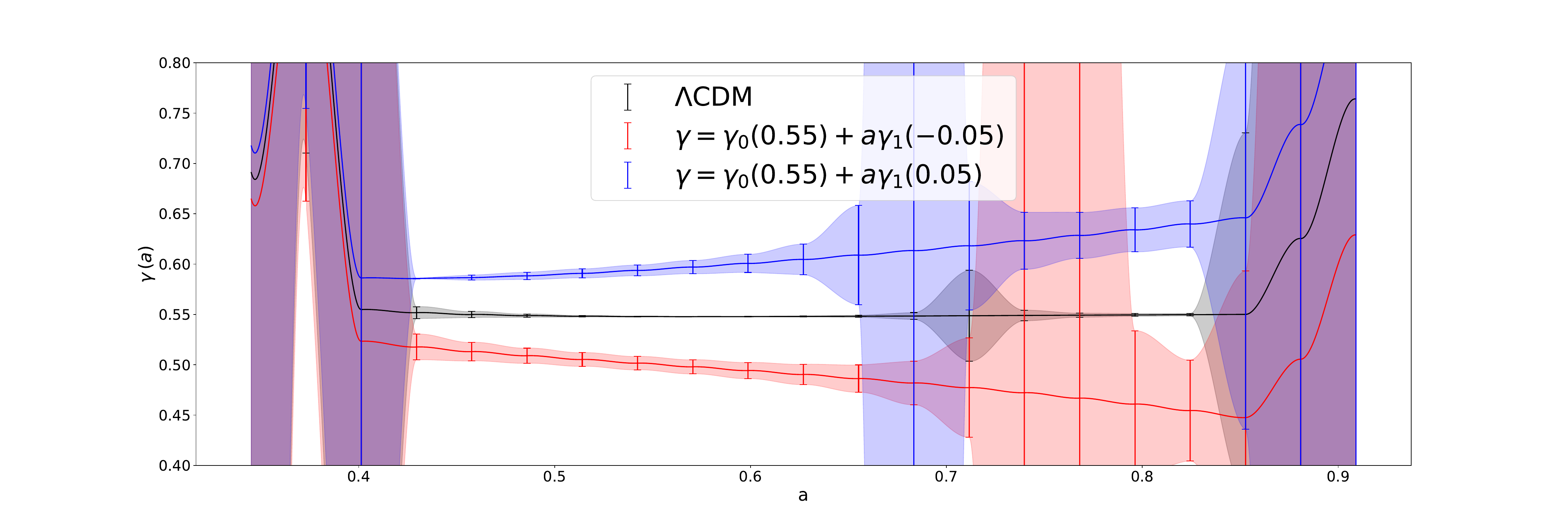}
\includegraphics[width=\textwidth]{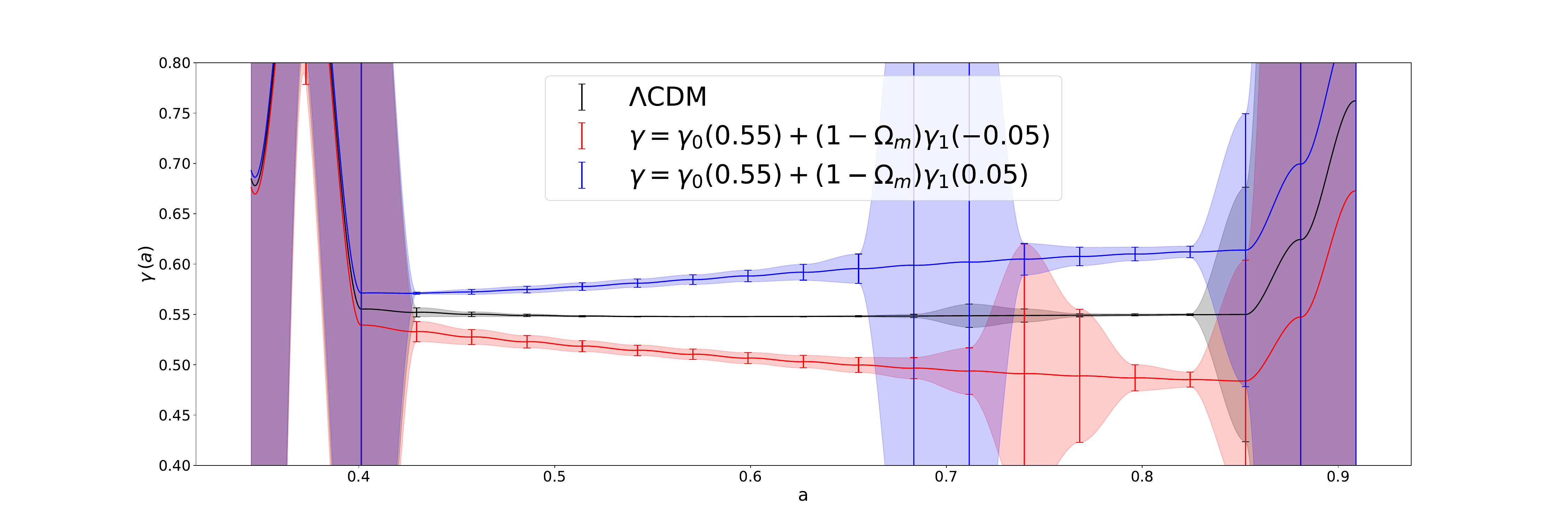}
\caption{Top panel: Forecast on $\gamma(a)$ supposing different observed values for $H(a)$, $f(a)$ and $f\sigma_8(a)$ following different asymptotic constant $\gamma$ values suitable for describing the shown modified gravity models (see Sect.~\ref{sect:TestandForcast} for details). Note that nDGP$_{\Lambda}$ signifies that we tuned it to have a $\Lambda$CDM background expansion. Middle and bottom panel:  same as above but for different parametrisations of $\gamma$ with a scale factor or $\Omega_{\rm DE}$ variation through an additional $\gamma_1$ parameter.}
\label{fig:GIgampure}
\end{figure*}

We start by showing in Fig.~\ref{fig:GIgampure} output of $\gamma(a)$ for different mock data, supposing the fiducial is either built following a $\gamma$ value for some common modified gravity models, or following a parametrisation with an additional $\gamma_1$ to $\gamma_0$ as function of the redshift or scale factor $a$ or function of the dark energy density or equivalently in flat space $1-\Omega_{\rm m}(a)$ \cite{Gong:2008fh}. The values of $\gamma$ for the different MG models are obtained with different considerations for each model. For DGP model we can determine its asymptotic value by a direct derivation as in \cite{LinderCahn2007}. This is allowed since this model can self accelerate, therefore its specific parameter $\Omega_{rc} = 1/4r_c^2H_0^2$ can be remapped into $\Omega_{\rm m}$ and $\gamma$. This also can be still done for nDGP providing we impose that the background behaves like $\Lambda$CDM where we reach values of 0.55-0.63 as weakly dependent of variation of $\Omega_{rc} $ \cite{Tsedrik:2024cdi}. For $f(R)$ or JBD, we have for the former to suppose that, first the background is $\Lambda$CDM and that we are working on small scales where it converges to values around 0.4-0.45 regardless or weakly dependent of the values of its specific parameter $f_{R0}$ \cite{Gannouji:2008wt}, while for the latter we needed to fix its parameter $w_{\rm JBD}$ to $\sim10$, a value that allows for a distinctive deviation from $\gamma$($\Lambda$CDM) to obtain however a small deviation with $\gamma \sim 0.565$. 

We see that all models in these assumptions could be ruled out if the data is $\Lambda$CDM with JBD the one that shows the least distinction followed by nDGP, while the normal branch DGP was already known to be in difficult situation even with current data.  $f(R)$ is also largely ruled out in the case of the assumptions we considered. We shall see later however, when we show more elaborate determination of $\gamma(a)$ in some MG frameworks,  different fates for $\gamma(a)$ for $f(R)$ model, and that following different values for its characteristic parameter $f_{R0}$.

\begin{figure*}[htbp]
\centering
\includegraphics[width=\textwidth]{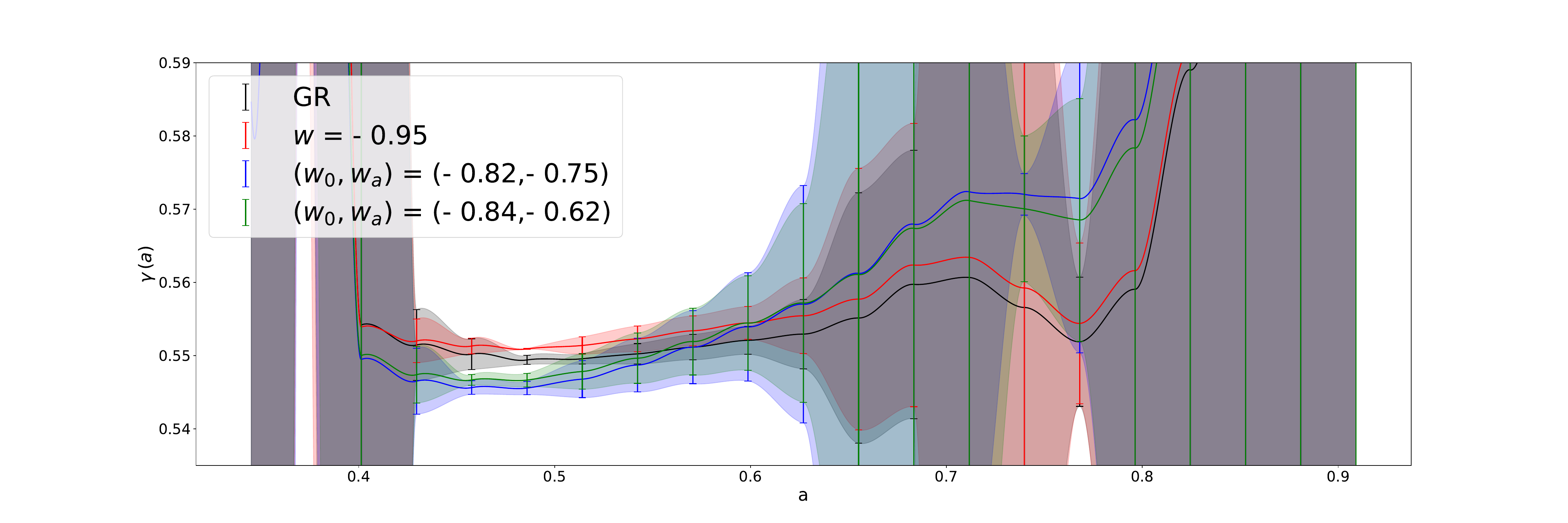}
\includegraphics[width=\textwidth]{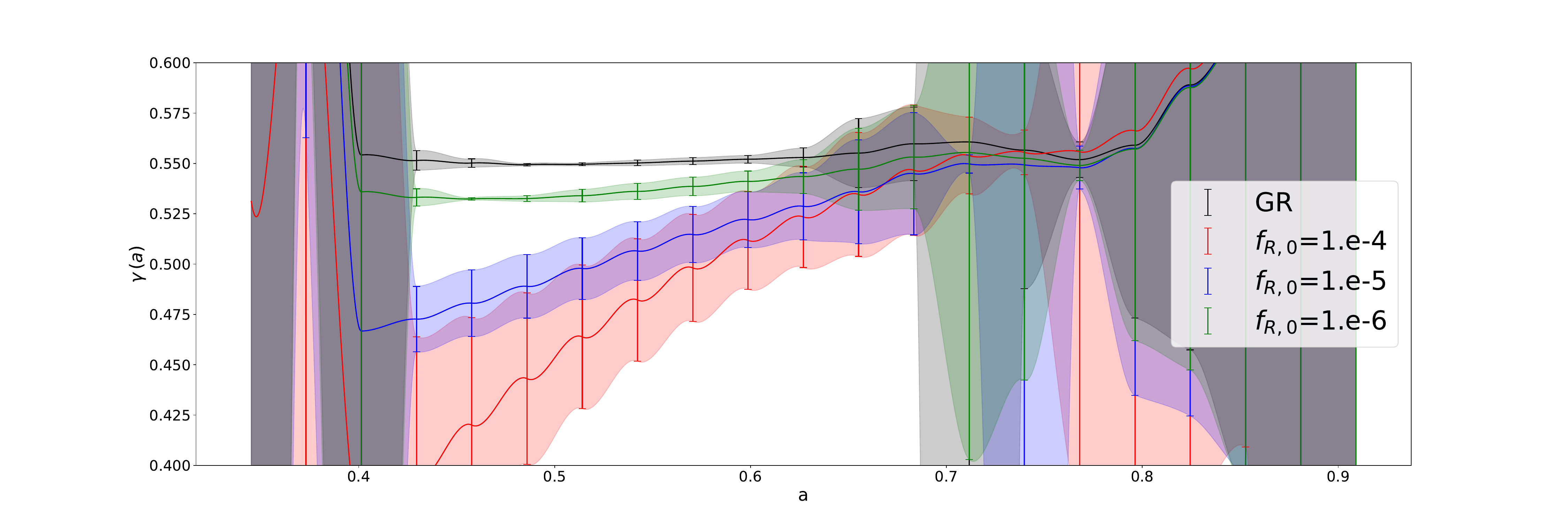}
\includegraphics[width=\textwidth]{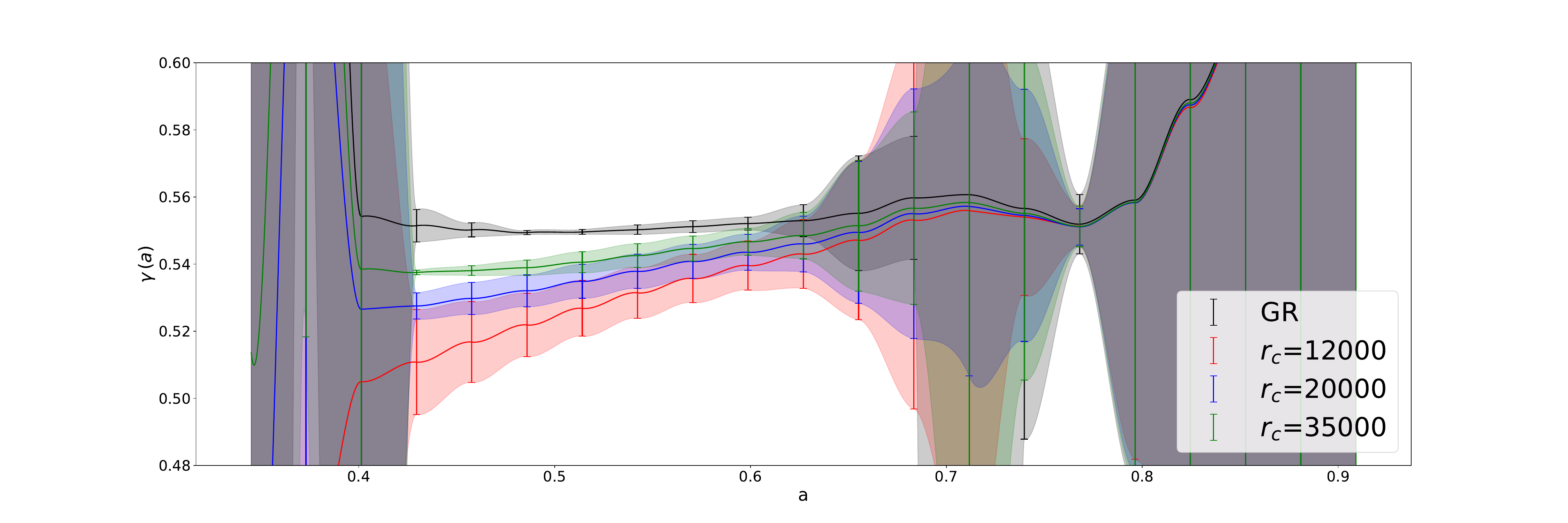}
\caption{Forecast on $\gamma(a)$ supposing different observed values for $H(a)$, $f(a)$ and $f\sigma_8(a)$ following different modified gravity model. Top panel: $w$ constant or CPL  dynamical dark energy models. Middle panel: $f(R)$ model. Bottom panel: nDGP model.}
\label{fig:GIgamMG}
\end{figure*}

The stringent errors we find supposing different $\gamma$ for different models enlarge in the case when we additionally extend to dynamical $\gamma(a)$ so that, either when $\gamma_1$ scales $a$ or $1-\Omega_m$, we are only able of disentangling from  $\Lambda$CDM case if we go to high redshifts with in general $\gamma_1 (1-\Omega_m)$ model showing smaller errors than when function of $a$ for the same fiducial value in each parameterisation. Note that the value of 0.05 adopted for $\gamma_1$ is within the values different studies find when they additionally allow an extension to $\gamma$ constant. The increase of the error bars is due to the fact that we propagate as function of the derivative of $\gamma(a)$ which for late time models as is the case here would inflate the errors for high values of $a$.

\begin{figure*}[htbp]
\centering
\includegraphics[width=\textwidth]{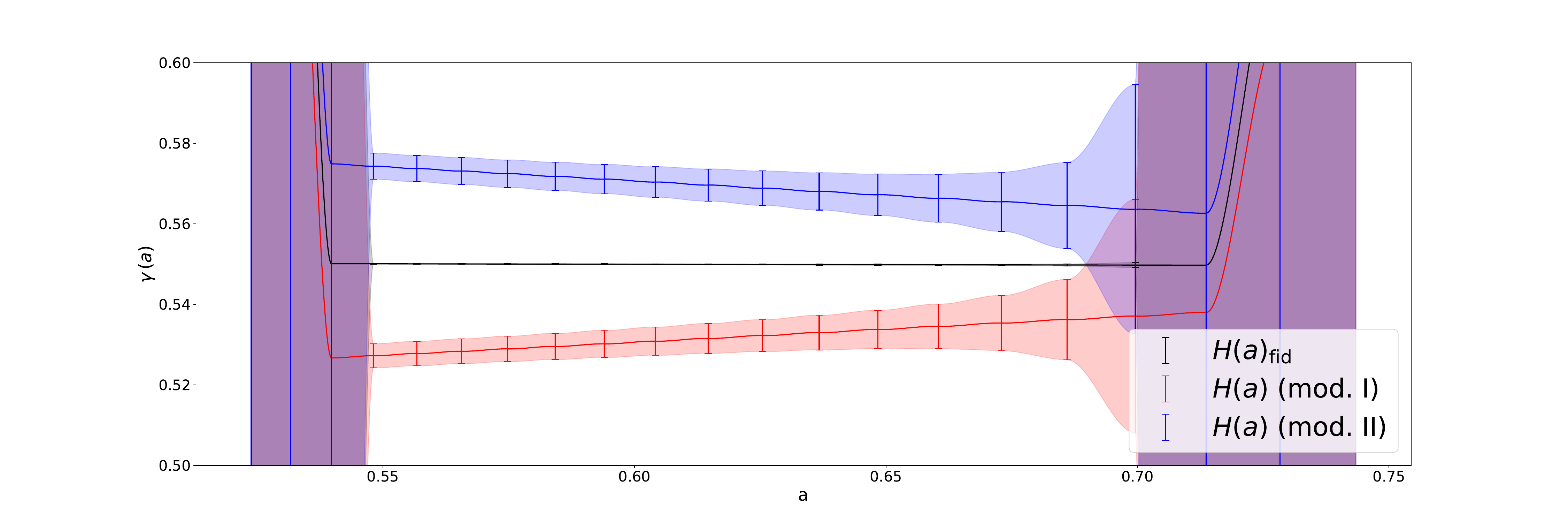}
\includegraphics[width=\textwidth]{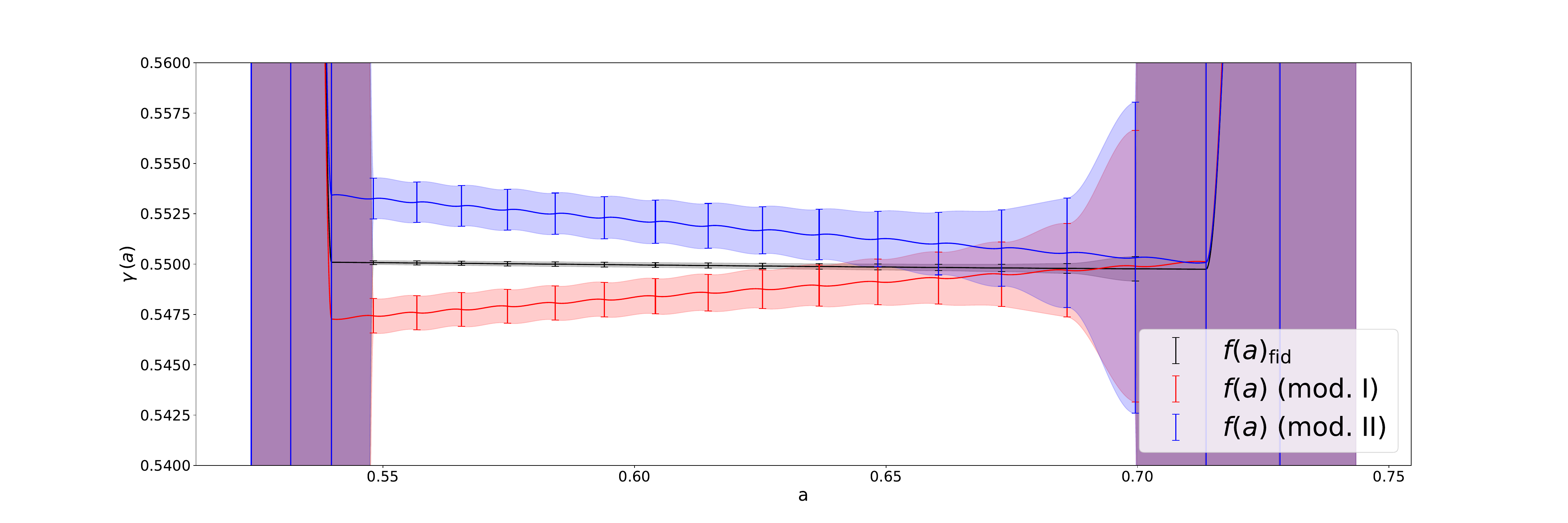}
\includegraphics[width=\textwidth]{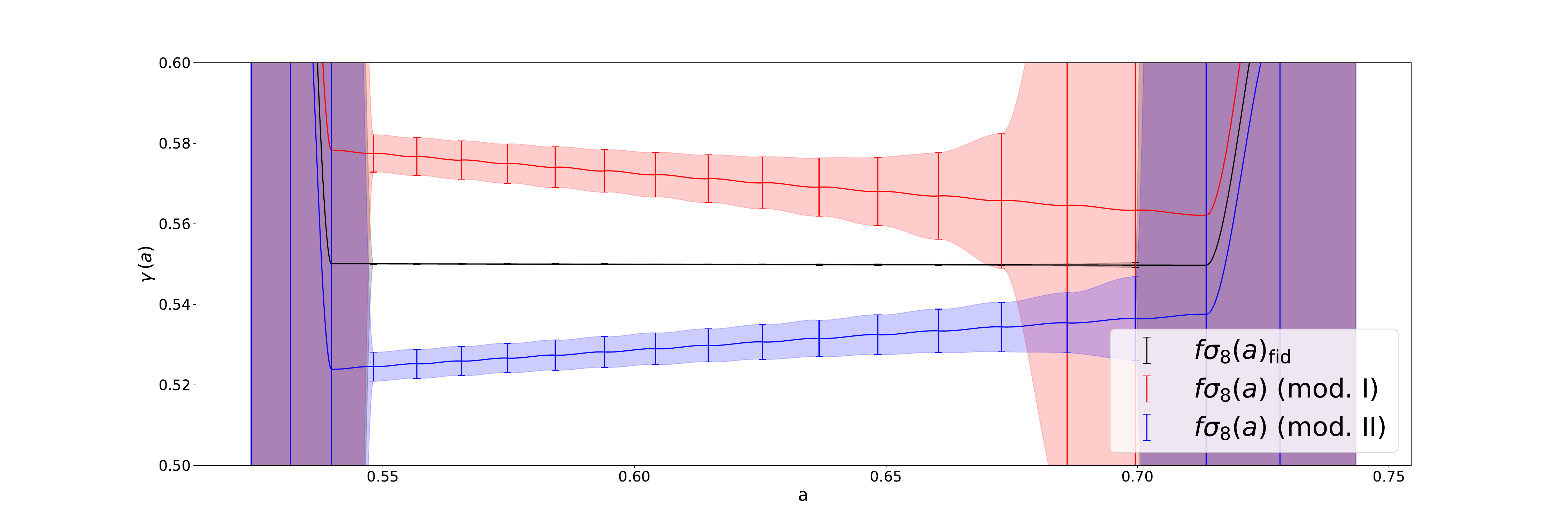}
\caption{Top panel: Forecast on $\gamma(a)$ supposing three different observed values for $H(a)$ while $f(a)$ and $f\sigma_8(a)$ follow the fiducial model. Middle panel:  Same as above but here $f(a)$ is being discrepant. Bottom panel: Same as above but supposing different $f\sigma_8(a)$  observations from different probes.}
\label{fig:GIgamdiscrep}
\end{figure*}

We next show $\gamma(a)$ supposing $H(a_i)$, $f(a_i)$ and $f\sigma_8 (a_i)$  are now produced by a cosmic solver such as $\texttt{MGCLASS}$ for three beyond $\Lambda$CDM models. We see first in Fig.~\ref{fig:GIgamMG} in the top panel $\gamma$ for a dynamical dark energy model, whether $w$ alone or $w_0w_a$ parameterisation with fiducial values being the best fit obtained from DESI I \cite{DESI:2024mwx} and II \cite{DESI:2025zgx} releases. These models will be very difficult to distinguish from $\Lambda$CDM even with Stage-IV surveys since $\gamma$ from the $w$ constant model is systematically higher than  $\gamma$($\Lambda$CDM) but well within the errors in all redshift bins, while $\gamma$ from $w_0w_a$ crosses that of $\Lambda$CDM and they can only be distinguished at high redshift. The situation is different for either $f(R)$ or nDGP in the middle and bottom panel for which we have chosen fiducial values close to current constraints from available data, where we see that they will be ruled out, if $\Lambda$CDM is the true model at almost all bins. We also that we can distinguish between the two extreme values for each model i.e.  $f_{R0}$ $\sim$ 10$^{-4}$ and 10$^{-6}$ for $f(R)$ and $r_c$ $\sim$ 12000 and 35000 for nDGP.

Finally, one of the purpose of our derivation, is to be able to catch possible discrepant measurments from different probes and surveys since each could be measured independently such as $H(a)$ from BAO or SN for example, $f\sigma_8$ from spectroscopic galaxy clustering surveys, while $f$ could simply reflect and be derived from a measurement of growth which could be obtained from photometric galaxy clustering or lensing angular correlation functions or galaxy clusters probes. As so we show in Fig.~\ref{fig:GIgamdiscrep} in the first upper panel $\gamma(a)$ from different observations  of $H(a)$ where the fiducial would be for an $\Omega_m \sim 0.315$ compatible with Planck 2018 \cite{Planck:2019nip} release best fit, while model I is for $\Omega_m \sim 0.33$ higher and compatible with DESI  \cite{DESI:2024mwx} or Pantheon+ \cite{Brout:2022vxf} data, and model II for $\Omega_m \sim 0.3$ close to the value needed to reconcile $\sigma_8$ from CMB with that found from lensing \cite{DES:2021wwk} or galaxy clusters probe \cite{Sakr:2018new}. We also show in the middle panel similar but for different $f(a)$ and then different $f\sigma_8(a)$ in the lower one. We observe that in all cases we will be able to disentangle between different values, with $f$ being the less affecting the shape of $\gamma(a)$, while $f\sigma_8$ impacts the most. Besides the nature of the component and how it changes the resulting $\gamma$ model, one other reason may be due to the fact that $f$ appear only in the derivation of $\gamma$ while $H$ enters also as the logarithm of its higher derivatives and $f\sigma_8$ enters through its first and second derivatives. We also see when we vary $f(a)$ that, as expected following Eq.~\ref{eq:gam}, that a higher $f$ implies a lower $\gamma$ and vice versa, while when we vary $H(a)$, we have to look at Eq.~\ref{eq:devgam} to get some insights: here $\gamma$ is constant and $f$, $f'$ are fixed so that for Mod I where we have a decrease in the expansion, $\gamma$ should go higher to adjust the equation. The situation for $f\sigma_8$ is less straightforward to explain given that the equations involving this term are not as simple to interpret unless it happens, as seen in Eq.~\ref{eq:gamma_code} that $\gamma$ is proportional to a dominant  $f\sigma_8$ in the denominator. But in general,  the behavior we see in the bottom panel is compatible with what we usually find from inference methods such as MCMC runs using current probes, where a discrepancy on a lower $\sigma_8$ as in Mod II is fixed by a higher growth or lower $\gamma$ \cite{Ilic:2019pwq}, hence as we observe, $f\sigma_8$ in Mod II is proportional to a high $\gamma$. 

\section{Conclusion}\label{sect:concl}

In this work we expressed the growth index $\gamma$, a key parameter which deviation from a fiducial value could signifie the need for models beyond $\Lambda$CDM, as function of observed quantities: the large scale structures growth rate $f(z)$, its product with the matter fluctuation $f\sigma_8(z)$ and the expansion factor $H(z)$. We were able then to estimate it at nodal points $z_i$ where the measurements are obtained. We then used the shape function method based on the conservation of the virtual energy principle, to show its connection with function interpolation methods, and construct a continuous function $\gamma(z)$. We also apply it, after propagating the errors at each nodal points $z_i$  to estimate the errors at each $z$. 

We then used the outcome to forecast the ability of Stage-IV surveys, such as Euclid and DESI, to disentangle between different parameterisations  of $\gamma$; or different common models beyond $\Lambda$CDM whether based on dynamical dark energy or modified gravity such as $f(R)$ and nDGP; or the possibility of a discrepancy between the three measurements. We find that commonly reported discrepancies between the measurements could be detected with our method, providing that the errors and lost of precision from our addition of degrees of freedom is small with respect to the resulting deviations of $\gamma$. The same could be concluded for first order extensions to $\gamma$ or the modified gravity models. The detection is however found to a lesser degree with models with dynamical dark energy when supposing the latest DESI reported values. We conclude that this method is a strong tool to investigate cosmology in a model-independent way especially with forthcoming data delivered with stringent uncertainties that would remedy for the decrease of constraining power from the increase of the degrees of freedom.

\bibliographystyle{cas-model2-names}
\bibliography{cas-refs}

@article{Ilic:2019pwq,
    author = "Ili{\'c}, St{\'e}phane and Sakr, Ziad and Blanchard, Alain",
    title = "{Cluster counts. II. Tensions, massive neutrinos, and modified gravity}",
    eprint = "1908.00163",
    archivePrefix = "arXiv",
    primaryClass = "astro-ph.CO",
    doi = "10.1051/0004-6361/201936423",
    journal = "Astron. Astrophys.",
    volume = "631",
    pages = "A96",
    year = "2019"
}

@Inbook{Pierron2012,
author="Pierron, Fabrice
and Gr{\'e}diac, Michel",
title="The Principle of Virtual Work",
bookTitle="The Virtual Fields Method: Extracting Constitutive Mechanical Parameters from Full-field Deformation Measurements",
year="2012",
publisher="Springer New York",
address="New York, NY",
pages="21--56",
isbn="978-1-4614-1824-5",
doi="10.1007/978-1-4614-1824-5_2",
url="https://doi.org/10.1007/978-1-4614-1824-5_2"
}

@article{Oliveira:2023uid,
    author = "Oliveira, Fernanda and Avila, Felipe and Bernui, Armando and Bonilla, Alexander and Nunes, Rafael C.",
    title = "{Reconstructing the growth index $\gamma $ with Gaussian processes}",
    eprint = "2311.14216",
    archivePrefix = "arXiv",
    primaryClass = "astro-ph.CO",
    doi = "10.1140/epjc/s10052-024-12953-w",
    journal = "Eur. Phys. J. C",
    volume = "84",
    number = "6",
    pages = "636",
    year = "2024"
}

@article{Arjona:2020kco,
    author = "Arjona, Rub{\'e}n and Nesseris, Savvas",
    title = "{Hints of dark energy anisotropic stress using Machine Learning}",
    eprint = "2001.11420",
    archivePrefix = "arXiv",
    primaryClass = "astro-ph.CO",
    reportNumber = "IFT-UAM/CSIC-20-18",
    doi = "10.1088/1475-7516/2020/11/042",
    journal = "JCAP",
    volume = "11",
    pages = "042",
    year = "2020"
}

@article{Yin:2019rgm,
    author = "Yin, Zhao-Yu and Wei, Hao",
    title = "{Observational Constraints on Growth Index with Cosmography}",
    eprint = "1902.00289",
    archivePrefix = "arXiv",
    primaryClass = "astro-ph.CO",
    doi = "10.1140/epjc/s10052-019-7191-8",
    journal = "Eur. Phys. J. C",
    volume = "79",
    number = "8",
    pages = "698",
    year = "2019"
}

@article{Perenon:2022fgw,
    author = "Perenon, Louis and Martinelli, Matteo and Maartens, Roy and Camera, Stefano and Clarkson, Chris",
    title = "{Measuring dark energy with expansion and growth}",
    eprint = "2206.12375",
    archivePrefix = "arXiv",
    primaryClass = "astro-ph.CO",
    doi = "10.1016/j.dark.2022.101119",
    journal = "Phys. Dark Univ.",
    volume = "37",
    pages = "101119",
    year = "2022"
}

@article{Lahav:1991wc,
    author = "Lahav, Ofer and Lilje, Per B. and Primack, Joel R. and Rees, Martin J.",
    title = "{Dynamical effects of the cosmological constant}",
    reportNumber = "PRINT-91-0084 (UC,SANTA-CRUZ)",
    journal = "Mon. Not. Roy. Astron. Soc.",
    volume = "251",
    pages = "128--136",
    year = "1991"
}

@article{Cortes:2024yon,
    author = "Cort{\^e}s, {\'I}caro B. S. and Batista, Ronaldo C.",
    title = "{Can dark energy explain a high growth index?}",
    eprint = "2411.00963",
    archivePrefix = "arXiv",
    primaryClass = "astro-ph.CO",
    doi = "10.1103/71jl-gmrn",
    journal = "Phys. Rev. D",
    volume = "112",
    number = "4",
    pages = "043532",
    year = "2025"
}

@article{Sakr:2023xnw,
    author = "Sakr, Ziad",
    title = "{A trium test on beyond $\Lambda$CDM triggering parameters}",
    eprint = "2305.02817",
    archivePrefix = "arXiv",
    primaryClass = "astro-ph.CO",
    doi = "10.1088/1475-7516/2023/08/080",
    journal = "JCAP",
    volume = "08",
    pages = "080",
    year = "2023"
}

@article{Sakr:2018new,
    author = "Sakr, Ziad and Ili{\'c}, St{\'e}phane and Blanchard, Alain and Bittar, Jamal and Farah, Wehbeh",
    title = "{Cluster counts: Calibration issue or new physics?}",
    eprint = "1803.11170",
    archivePrefix = "arXiv",
    primaryClass = "astro-ph.CO",
    doi = "10.1051/0004-6361/201833151",
    journal = "Astron. Astrophys.",
    volume = "620",
    pages = "A78",
    year = "2018"
}

@article{DES:2021wwk,
    author = "Abbott, T. M. C. and others",
    collaboration = "DES",
    title = "{Dark Energy Survey Year 3 results: Cosmological constraints from galaxy clustering and weak lensing}",
    eprint = "2105.13549",
    archivePrefix = "arXiv",
    primaryClass = "astro-ph.CO",
    reportNumber = "FERMILAB-PUB-21-221-AE, DES-2020-0617",
    doi = "10.1103/PhysRevD.105.023520",
    journal = "Phys. Rev. D",
    volume = "105",
    number = "2",
    pages = "023520",
    year = "2022"
}

@article{Planck:2019nip,
    author = "Aghanim, N. and others",
    collaboration = "Planck",
    title = "{Planck 2018 results. V. CMB power spectra and likelihoods}",
    eprint = "1907.12875",
    archivePrefix = "arXiv",
    primaryClass = "astro-ph.CO",
    doi = "10.1051/0004-6361/201936386",
    journal = "Astron. Astrophys.",
    volume = "641",
    pages = "A5",
    year = "2020"
}

@article{Brout:2022vxf,
    author = "Brout, Dillon and others",
    title = "{The Pantheon+ Analysis: Cosmological Constraints}",
    eprint = "2202.04077",
    archivePrefix = "arXiv",
    primaryClass = "astro-ph.CO",
    doi = "10.3847/1538-4357/ac8e04",
    journal = "Astrophys. J.",
    volume = "938",
    number = "2",
    pages = "110",
    year = "2022"
}

@article{Sakr:2023bms,
    author = "Sakr, Ziad",
    title = "{Untying the Growth Index to Relieve the {\ensuremath{\sigma}}$_{8}$ Discomfort}",
    eprint = "2305.02863",
    archivePrefix = "arXiv",
    primaryClass = "astro-ph.CO",
    doi = "10.3390/universe9080366",
    journal = "Universe",
    volume = "9",
    number = "8",
    pages = "366",
    year = "2023"
}

@article{Tsedrik:2024cdi,
    author = "Tsedrik, Maria and Bose, Benjamin and Carrilho, Pedro and Pourtsidou, Alkistis and Pamuk, Sefa and Casas, Santiago and Lesgourgues, Julien",
    title = "{Stage-IV cosmic shear with Modified Gravity and model-independent screening}",
    eprint = "2404.11508",
    archivePrefix = "arXiv",
    primaryClass = "astro-ph.CO",
    doi = "10.1088/1475-7516/2024/10/099",
    journal = "JCAP",
    volume = "10",
    pages = "099",
    year = "2024"
}

@article{Gong:2008fh,
    author = "Gong, Yungui",
    title = "{The growth factor parameterization and modified gravity}",
    eprint = "0808.1316",
    archivePrefix = "arXiv",
    primaryClass = "astro-ph",
    doi = "10.1103/PhysRevD.78.123010",
    journal = "Phys. Rev. D",
    volume = "78",
    pages = "123010",
    year = "2008"
}

@article{Gannouji:2008wt,
    author = "Gannouji, R. and Moraes, B. and Polarski, D.",
    title = "{The growth of matter perturbations in f(R) models}",
    eprint = "0809.3374",
    archivePrefix = "arXiv",
    primaryClass = "astro-ph",
    doi = "10.1088/1475-7516/2009/02/034",
    journal = "JCAP",
    volume = "02",
    pages = "034",
    year = "2009"
}

@article{Sakr:2021ylx,
    author = "Sakr, Ziad and Martinelli, Matteo",
    title = "{Cosmological constraints on sub-horizon scales modified gravity theories with MGCLASS II}",
    eprint = "2112.14175",
    archivePrefix = "arXiv",
    primaryClass = "astro-ph.CO",
    doi = "10.1088/1475-7516/2022/05/030",
    journal = "JCAP",
    volume = "05",
    number = "05",
    pages = "030",
    year = "2022"
}

@article{Zheng:2023yco,
    author = "Zheng, Ziyang and Sakr, Ziad and Amendola, Luca",
    title = "{Testing the cosmological Poisson equation in a model-independent way}",
    eprint = "2312.07436",
    archivePrefix = "arXiv",
    primaryClass = "astro-ph.CO",
    doi = "10.1016/j.physletb.2024.138647",
    journal = "Phys. Lett. B",
    volume = "853",
    pages = "138647",
    year = "2024"
}

@article{Amendola:2023awr,
    author = "Amendola, Luca and Marinucci, Marco and Pietroni, Massimo and Quartin, Miguel",
    title = "{Improving precision and accuracy in cosmology with model-independent spectrum and bispectrum}",
    eprint = "2307.02117",
    archivePrefix = "arXiv",
    primaryClass = "astro-ph.CO",
    doi = "10.1088/1475-7516/2024/01/001",
    journal = "JCAP",
    volume = "01",
    pages = "001",
    year = "2024"
}

@article{Wang1998,
author = {Wang, L. and Steinhardt, P. J.},
title = {Cluster abundance constraints on quintessence models},
journal = {Astrophys. J.},
volume = {508},
pages = {483--490},
year = {1998},
url = {https://ui.adsabs.harvard.edu/abs/1998ApJ...508..483W}
}

@article{Linder2005,
author = {Linder, E. V.},
title = {Cosmic growth history and expansion history},
journal = {Phys. Rev. D},
volume = {72},
pages = {043529},
year = {2005},
doi = {10.1103/PhysRevD.72.043529},
url = {https://doi.org/10.1103/PhysRevD.72.043529}
}

@article{Specogna2024,
author = {Specogna, Enrico and Di Valentino, Eleonora and Said, Jackson Levi and Nguyen, Nhat-Minh},
title = {Exploring the growth index},
journal = {Phys. Rev. D},
volume = {109},
pages = {043528},
year = {2024},
doi = {10.1103/PhysRevD.109.043528},
url = {https://doi.org/10.1103/PhysRevD.109.043528}
}

@article{LinderCahn2007,
author = {Linder, E. V. and Cahn, R. N.},
title = {Parameterized beyond-Einstein growth},
journal = {Astropart. Phys.},
volume = {28},
pages = {481--488},
year = {2007},
doi = {10.1016/j.astropartphys.2007.09.004},
url = {https://doi.org/10.1016/j.astropartphys.2007.09.004}
}

@article{Pogosian2010,
author = {Pogosian, L. and Silvestri, A. and Koyama, K. and Zhao, G.-B.},
title = {How to optimally parametrize deviations from General Relativity in the evolution of cosmological perturbations},
journal = {Phys. Rev. D},
volume = {81},
pages = {104023},
year = {2010},
doi = {10.1103/PhysRevD.81.104023},
url = {https://doi.org/10.1103/PhysRevD.81.104023}
}

@article{Polarski2008,
author = {Polarski, D. and Gannouji, R.},
title = {On the growth of linear perturbations},
journal = {Phys. Lett. B},
volume = {660},
pages = {439--443},
year = {2008},
doi = {10.1016/j.physletb.2008.01.061},
url = {doi.org/10.1016/j.physletb/2008.01.061}
}

@article{Basilakos:2016nyg,
    author = "Basilakos, Spyros and Nesseris, Savvas",
    title = "{Testing Einstein{\textquoteright}s gravity and dark energy with growth of matter perturbations: Indications for new physics?}",
    eprint = "1610.00160",
    archivePrefix = "arXiv",
    primaryClass = "astro-ph.CO",
    reportNumber = "IFT-UAM-CSIC-16-095",
    doi = "10.1103/PhysRevD.94.123525",
    journal = "Phys. Rev. D",
    volume = "94",
    number = "12",
    pages = "123525",
    year = "2016"
}

@article{Steigerwald:2014ava,
    author = "Steigerwald, Heinrich and Bel, Julien and Marinoni, Christian",
    title = "{Probing non-standard gravity with the growth index: a background independent analysis}",
    eprint = "1403.0898",
    archivePrefix = "arXiv",
    primaryClass = "astro-ph.CO",
    doi = "10.1088/1475-7516/2014/05/042",
    journal = "JCAP",
    volume = "05",
    pages = "042",
    year = "2014"
}

@article{Wen:2023bcj,
    author = "Wen, Yuewei and Nguyen, Nhat-Minh and Huterer, Dragan",
    title = "{Sweeping Horndeski canvas: new growth-rate parameterization for modified-gravity theories}",
    eprint = "2304.07281",
    archivePrefix = "arXiv",
    primaryClass = "astro-ph.CO",
    reportNumber = "LCTP-23-05",
    doi = "10.1088/1475-7516/2023/09/028",
    journal = "JCAP",
    volume = "09",
    pages = "028",
    year = "2023"
}

@article{DESI:2024mwx,
    author = "Adame, A. G. and others",
    collaboration = "DESI",
    title = "{DESI 2024 VI: cosmological constraints from the measurements of baryon acoustic oscillations}",
    eprint = "2404.03002",
    archivePrefix = "arXiv",
    primaryClass = "astro-ph.CO",
    reportNumber = "FERMILAB-PUB-24-0154-PPD",
    doi = "10.1088/1475-7516/2025/02/021",
    journal = "JCAP",
    volume = "02",
    pages = "021",
    year = "2025"
}

@article{DESI:2025zgx,
    author = "Abdul Karim, M. and others",
    collaboration = "DESI",
    title = "{DESI DR2 results. II. Measurements of baryon acoustic oscillations and cosmological constraints}",
    eprint = "2503.14738",
    archivePrefix = "arXiv",
    primaryClass = "astro-ph.CO",
    reportNumber = "FERMILAB-PUB-25-0169-PPD",
    doi = "10.1103/tr6y-kpc6",
    journal = "Phys. Rev. D",
    volume = "112",
    number = "8",
    pages = "083515",
    year = "2025"
}

@article{Euclid:2024yrr,
    author = "Mellier, Y. and others",
    collaboration = "Euclid",
    title = "{Euclid. I. Overview of the Euclid mission}",
    eprint = "2405.13491",
    archivePrefix = "arXiv",
    primaryClass = "astro-ph.CO",
    doi = "10.1051/0004-6361/202450810",
    journal = "Astron. Astrophys.",
    volume = "697",
    pages = "A1",
    year = "2025"
}

@article{LSSTScience:2009jmu,
    author = "Abell, Paul A. and others",
    collaboration = "LSST Science, LSST Project",
    title = "{LSST Science Book, Version 2.0}",
    eprint = "0912.0201",
    archivePrefix = "arXiv",
    primaryClass = "astro-ph.IM",
    reportNumber = "FERMILAB-TM-2495-A, SLAC-R-1031",
    doi = "10.2172/1156415",
    month = "12",
    year = "2009"
}

@article{DESI:2016igz,
    author = "Aghamousa, Amir and others",
    collaboration = "DESI",
    title = "{The DESI Experiment Part II: Instrument Design}",
    eprint = "1611.00037",
    archivePrefix = "arXiv",
    primaryClass = "astro-ph.IM",
    reportNumber = "FERMILAB-PUB-16-518-AE",
    month = "10",
    year = "2016"
}

\end{document}